\documentclass[preprint,aps]{revtex4}

\usepackage{graphicx}

\begin{document}

\title{Exact bound states in volcano potentials}   

\author{Ratna Koley \footnote{Electronic address: 
{\em ratna@cts.iitkgp.ernet.in}} ${}^{}$ 
and 
Sayan Kar \footnote{Electronic address : {\em sayan@cts.iitkgp.ernet.in}}${}^{}$}
\affiliation{Department of Physics and Centre for Theoretical Studies  \\
Indian Institute of Technology, Kharagpur 721 302, India}

\begin{abstract}
Quantum mechanics in a one--parameter family of volcano potentials is 
investigated. After a discussion on their construction and classical 
mechanics, we obtain exact, normalisable bound states for specific values 
of the energy. The nature of the wave functions and probability 
densities, as well as some curious features of the solutions are
highlighted. 
\end{abstract}

\maketitle

\noindent {\sf Introduction :}
Volcano potentials are not quite common in traditional quantum
mechanics. A generic potential of this kind has a depression (well) at the 
origin with its value approaching negative infinity 
asymptotically (on both sides of the origin). The dip at 
the origin with finite--height barriers on either side is the reason behind the
name {\em volcano} \cite{volcano, resonance}. One might have another type of 
{\em volcano potential} for which the asymptotics is different--it goes to 
zero, or, a constant value, asymptotically. The former type is the
inverted version of the well--known double well whereas the latter is
the so--called double barrier {\cite{semi}}.  
In this article, we focus on the former type of potentials, i.e. the ones
which go to negative infinity asymptotically. It must be noted that a
crucial difference between the two types is the fact that the former, 
by definition, has non--Hermitian boundary conditions \cite{resonance}. 

Recently, volcano potentials have arisen 
in various contexts. Both the types mentioned in the previous paragraph 
appear in high energy physics in the context of localization of spin zero, 
half and spin two fields in the currently popular braneworld models 
{\cite{braneworld}}. For example, in the five dimensional Randall--Sundrum 
model, the equation for the Kaluza--Klein modes of the graviton 
reduces to a Schr\"{o}dinger-like equation in a volcano potential of either 
type (using a coordinate transformation one can go from one type to the 
other, though this may not be possible always). On the other hand,
double barrier structures (volcano box potentials) arise in studies related 
to artificial quantum heterostructures (quantum wires, wells and dots) 
and are reasonably well--known {\cite{semi}}. 
It is therefore relevant to look at model volcano potentials
and try to understand standard quantum mechanics in their
presence. Several authors have made such attempts in the
recent past {\cite{volcano,resonance}}. Our aim here is to provide further
illustrations and insight for a specific one--parameter family of volcano 
potentials of the first type, primarily through a class of exact solutions.  

What do we expect quantum--mechanically, if, say, a quantum particle
feels such a potential ? The natural answer is that resonances
appear in the spectrum which correspond to complex eigenvalues. 
The resonant states have
finite masses and widths. They may tunnel out and disappear -- but 
using known formalism we can 
estimate how long they might exist and provide the illusion of a
possible bound state. Extensive work on understanding the resonances of
the volcano have been reported in {\cite{resonance,zamastil1}}.
On the other hand, it is also not {\em impossible}
to have bound states with a real spectrum. Examples such as the
well known $ax^2-bx^4$ potential have been studied quite extensively
in the context of the recently formulated PT symmetric quantum
mechanics {\cite{ptsymm}} where the non--Hermiticity of the Schrodinger
operator gives rise to distinct features not quite known in standard
quantum mechanics. 

\noindent {\sf  The volcano potentials:}
We begin by proposing a somewhat general construction of a volcano
potential using arbitrary
but well--behaved functions f(x) and h(x). 
One can obtain various profiles
for the potential choosing different forms of 
these two functions. In order to generate a volcano potential
we must have specific properties for f(x) and h(x). We delineate
these below for symmetric potentials.

(i) f(x) is an even function, is finite (may be zero) at the origin ($x=0$)
and increases monotonically (to positive infinity) as $x \rightarrow \pm \infty$.

(ii) h(x) is also an even function, is finite (may be zero) at the origin
($x=0$) but decreases monotonically (to negative infinity) as $x
\rightarrow \pm \infty$ 
at a rate faster than the increase of $f(x)$.

As an example, let us consider $f(x)=a x^2$ and $h(x)=-b x^4$, where $a>0$ and
$b>0$ are two constants. The resultant $V(x) = f(x)+h(x)$ is the
inverted double well and represents a volcano potential.
It is easy to construct many other examples.     

A couple of points about classical mechanics in this system. 

(i) A particle with energy $E<V_{max}$ and with initial position within
$-x_1 <x<x_1$ ($x_1, -x_1$ and $x_2, -x_2$ are the four points of
intersection of the $E=$ constant line with the potential, $x_1<x_2$  
and $E=V(\pm x_1, \pm x_2)$) will oscillate about $x=0$. It will
reside in the volcano and cannot come out. Within the barriers, i.e.
for initial conditions between $x_1 <x < x_2$ or $-x_1 > x > -x_2$ there are
no classical solutions. However if the initial position
is such that $ x > x_2$ or $x<-x_2$ then the particle will disappear to
positive or negative infinity. A nice discussion on the phase flow, fixed
points and invariant sets of a first order autonomous system in a 
generic volcano potential of the type
$x^2-x^4$ is available in {\cite{percival}}.

(ii) If $E=V_{max}$ then the particle resides at an unstable point and
depending on the initial velocity can either slide off to $\pm \infty$
or down into the volcano. The usual `Euclidean time' instanton solutions
for the double well will now be valid for `real' time. 

Another way to construct a family of potentials which includes the
volcano, is by using a single function
and its derivatives.
Let us write down 
an expression for such potentials using the following ansatz:

\begin{equation}  
V(x) = - (a_{1} e^{2 g(x)} + a_{2} g''(x))
\end{equation}
where, $a_{1}$ and $a_{2}$ are two arbitrary constants. 
We assume $g(x)$ to be such that $e^{2 g(x)}$ and $g''(x)$ are positive for all 
$x$.
For diverse choices of the function, g(x) one can obtain different potentials, 
each of which will have a definite shape (single barrier, volcano, single
well and double well). $g(x)$ therefore, is some kind of
generating function for the potentials. For example, let us
consider g(x) $ = \mbox{ln (sech x)}$.
In this case, the potential has single barrier and is proportional to 
$\mbox{sech}^2 x$. 
Similarly for g(x) = $\mbox{ln} (\cosh^{\nu} x)$, the general form of the 
potential becomes 

\begin{equation}  
V(x) = - (a_{1} \cosh^{2 \nu} x + a_{2} \mbox{sech}^2 x) \label{potential}
\end{equation} 
where, $\nu$ is an arbitrary constant. Now, for distinct 
values of the constants the potential takes various forms. Let us discuss 
some of the possibilities in more detail. 

\noindent {\em Case I: ( $a_{1}$ and $a_{2}$ both positive )}

\begin{figure}[htb]
\includegraphics[width= 8cm,height=5cm]{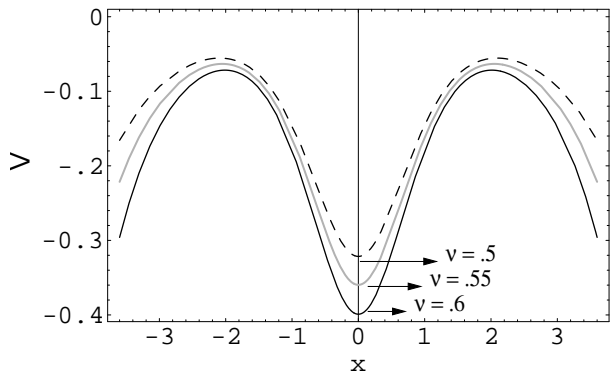}
\caption{The potential, 2V is plotted as a function of x for
 different values of the parameter $\nu$, 
for $a_{1}$ = $0.009$ and 
$a_{2}$ = $\frac{\nu}{2} (\frac{\nu}{2} + 1)$. The depth at $x=0$ 
increases with increasing $\nu$.}
\end{figure}

In this case one can obtain a double-barrier volcano potential with V(x) negative
for all values of x. It is possible to have bound states at the location of the well.
Assuming $a_{2} = \nu (\nu+1) $ we find a minimum at $x=0$ if
$a_2>\nu a_1$. The profile 
depends on the parameter $\nu$ 
[Fig. (1)]. For the particle energies lesser than the depth 
of the potential well, there will be bound
states. The potential decreases again beyond the turning points. Thus,
there is a finite probability of
tunneling through either barrier. This can be seen by assuming
radiative boundary conditions at either infinity and calculating the
tunneling probability in the same way as done for barrier penetration
problems. As a result, there could be
states found within the well which are quasibound. More precisely, these
would be solutions with complex eigenvalues -- the imaginary part being
related to the lifetime of the quasibound state in the 
well. Some recent methods for the calculation of the lifetime
of the quasibound states have been proposed in \cite{zamastil1}. 
We have plotted the potential for different
values of the parameter $\nu$. It is apparent from Fig. (1) that the 
depth of the potential well
increases with increasing values of the parameter $\nu$ ($a_2$). As a
consequence, the probability of having a large number of bound 
states within the well gets more and more pronounced with increasing $\nu$. 

\noindent {\em Case II : ( $a_{1}=0$ and $a_{2}$ positive )} 

The potential obtained in this case is the well--known 
P\"{o}sch Teller potential (Fig. 2). The solutions of the 
Schr\"{o}dinger equation reveals that the energy spectrum is continuous 
for the positive eigen values whereas it is discrete for negative 
eigenvalues of the energy \cite{lf}. Discrete bound states are found 
inside the well.  

\begin{figure}[h]
\includegraphics[width= 8cm,height=5cm]{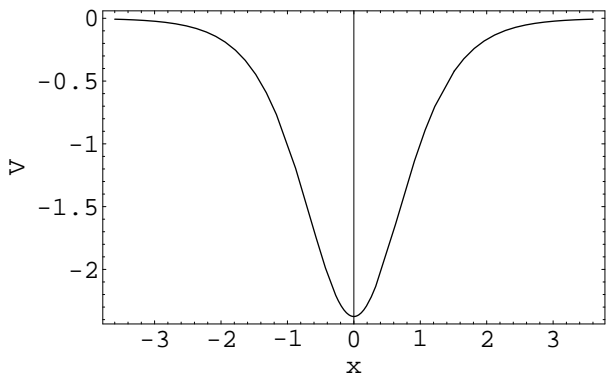}
\caption{The potential reduces to the well known  P\"{o}sch-Teller
potential for $a_{1} = 0$  and $a_{2}$ positive.}
\end{figure}

\noindent {\em Case III \&  IV : ( $a_{1}$, $a_{2}$ both negative /  
$a_{1}$ -ve, $a_{2}$ +ve )} 

For the single barrier potential (Case III) we essentially have the usual
barrier penetration problem. In the last case,
{\em{i.e.}} for a potential well, we can obtain bound states with a
discrete energy spectrum. We intend to focus on the double-barrier
volcano potential in the remaining part of this article. 

Before we go over to solving the Schrodinger equation let us note
the following fact. The volcano potential in the Case I above is
not entirely unrelated to the inverted double well which we discussed
at the beginning of this section. To see this, consider a Taylor
expansion of the terms in the $cosh$--$sech$ volcano and retain
terms up to order $x^4$. One will notice the reappearance of
the $a x^2-b x^4$ potential for $\nu a_1<a_2$. Polynomial approximations 
for the other cases (II, III, IV)
may also be obtained using the same expansion.  

\noindent {\sf Exact solutions of the Schr\"{o}dinger Equation:}
For the potential given in Eqn. (\ref{potential}), the Schr\"{o}dinger equation
is of the following form :

\begin{equation}
\frac{d^2 \psi (x)}{ d x^2} + (A_{1} \cosh^{2 \nu} x + A_{2}
\hspace{.1 cm} \mbox{sech}^2 x + A_{3}) \psi (x) = 0 \label{schequation}
\end{equation}  
where, $A_{1} = 2 a_{1},  A_{2} = 2 a_{2}$  and  $A_{3} = 2 E$. It is
possible to find an exact solution
of the above equation if the relations, $A_{2} =  \frac{\nu}{2}(
\frac{\nu}{2} +1)$ and $A_{3} = - \frac{\nu^2}{4}$ are 
satisfied. The most general solution to Eqn. (\ref{schequation}) 
is a combination of two linearly independent solutions,
$\psi_{1}(x)$ and $\psi_{2}(x)$, given by 

\begin{eqnarray}
\psi_{1}(x) = A \frac{\cos \left[\sqrt{A_{1}}  
\hspace{.1 cm} \int (\cosh x)^{\nu} dx \right]}{(\cosh x)^{\frac{\nu}{2}}} \\
\psi_{2}(x) = B \frac{\sin \left[\sqrt{A_{1}}  
\hspace{.1 cm} \int (\cosh x)^{\nu} dx \right]}{(\cosh x)^{\frac{\nu}{2}}} 
\end{eqnarray}
where, A and B are arbitrary constants, to be determined by the
normalization condition. It is manifest from the general
expressions that $\nu$ should be positive for the wave function to be
finite at x $\rightarrow$ $\pm \infty$. The wave function is normalizable with 
$A = \left( \frac{\pi}{2} (1 + e^{-2 \sqrt{A_{1}}}) \right)^{-\frac{1}{2}}$ and 
$B = \left( \frac{\pi}{2} (1 - e^{-2 \sqrt{A_{1}}})
\right)^{-\frac{1}{2}}$ for $\nu = 1$ with the normalization constant
chosen as unity. The total energy of the 
particle is negative and the motion of the particle is bounded between the classical 
turning 
points of the potential when the particle's energy lies between
$V_{min}$ 
and the peak value of the 
potential {\em {i.e.}} $V_{min} < E < V_{max}$. The fact that the value of the
bound state energy be within the well imposes a restriction of the allowed
domain of $A_1$. This turns out to be $A_1>\frac{\nu}{2}$.
Let us now explore how the states depend 
on different values of $\nu$.
We show this graphically. The
number of nodes of the wave function increases
with increasing $\nu$ within a given region. In Fig. (3), we have plotted the wave functions
for $\nu = 1$ and $\nu = 4$. The normalization constants are obtained
for $\nu = 1$ and one can argue that the wave functions are also normalizable for $\nu =
4$ in the sense that ${\vert \psi \vert}^2$ covers a finite area for x
$\rightarrow$ $\pm \infty$. 

\begin{figure}[h]
\includegraphics[width= 8cm,height=5cm]{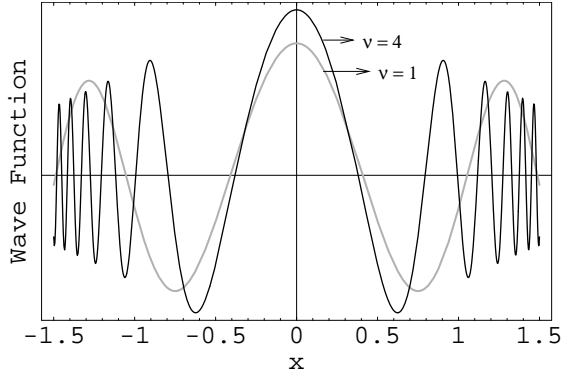}
\caption{The even parity wave function, $\psi_{1}$(x) is plotted as a function of x for
 two different values of the parameter $\nu$ = 1 and 4.
It is clear that the number of nodes increases with $\nu$ increasing.}
\end{figure}

\begin{figure}[h]
\includegraphics[width= 8cm,height=5cm]{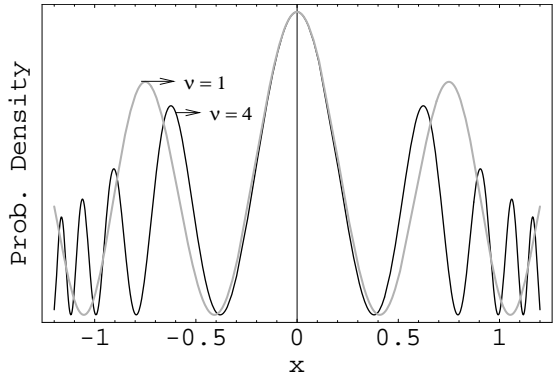}
\caption{The probability density is plotted as a function of x for
 two different values of the parameter $\nu$ = 1 and 4 for even states.
As both the curves are showing finite area we can conclude
 the wave functions are normalizable.}
\end{figure}
  
The probability density of the even states for
the same set of $\nu$ values have been depicted in Fig. (4). From this
we can conclude that the wave functions are normalizable and well
behaved. The nature of the wave functions and probability density for
the odd states with different values of
the parameter $\nu$ are
shown in Fig. (5) and (6) respectively.

\begin{figure}[h]
\includegraphics[width= 8cm,height=5cm]{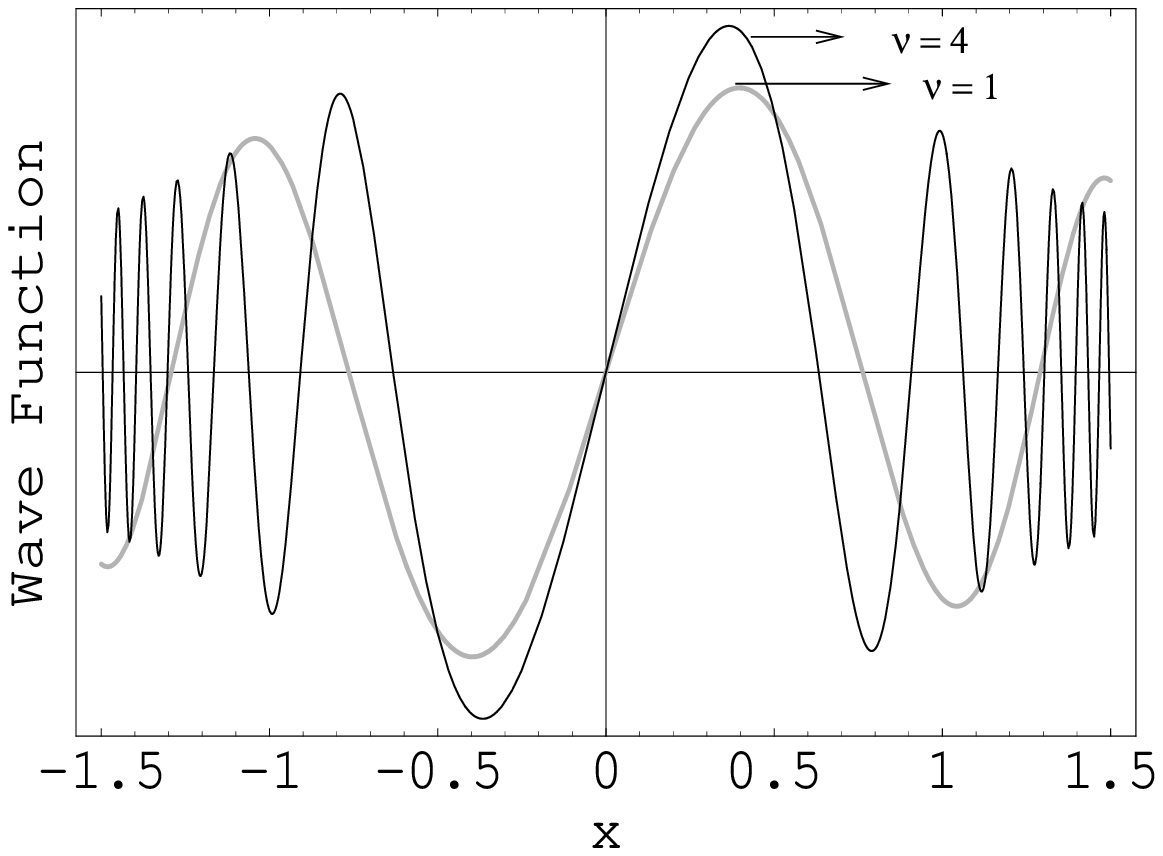}
\caption{The odd parity wave function, $\psi_{2}$(x) is plotted as a function of x for
 two different values of the parameter $\nu =$ 1 and 4.
It is clear that the number of nodes increases with $\nu$ increasing.}
\end{figure}

\begin{figure} [h]
\includegraphics[width= 8cm,height=5cm]{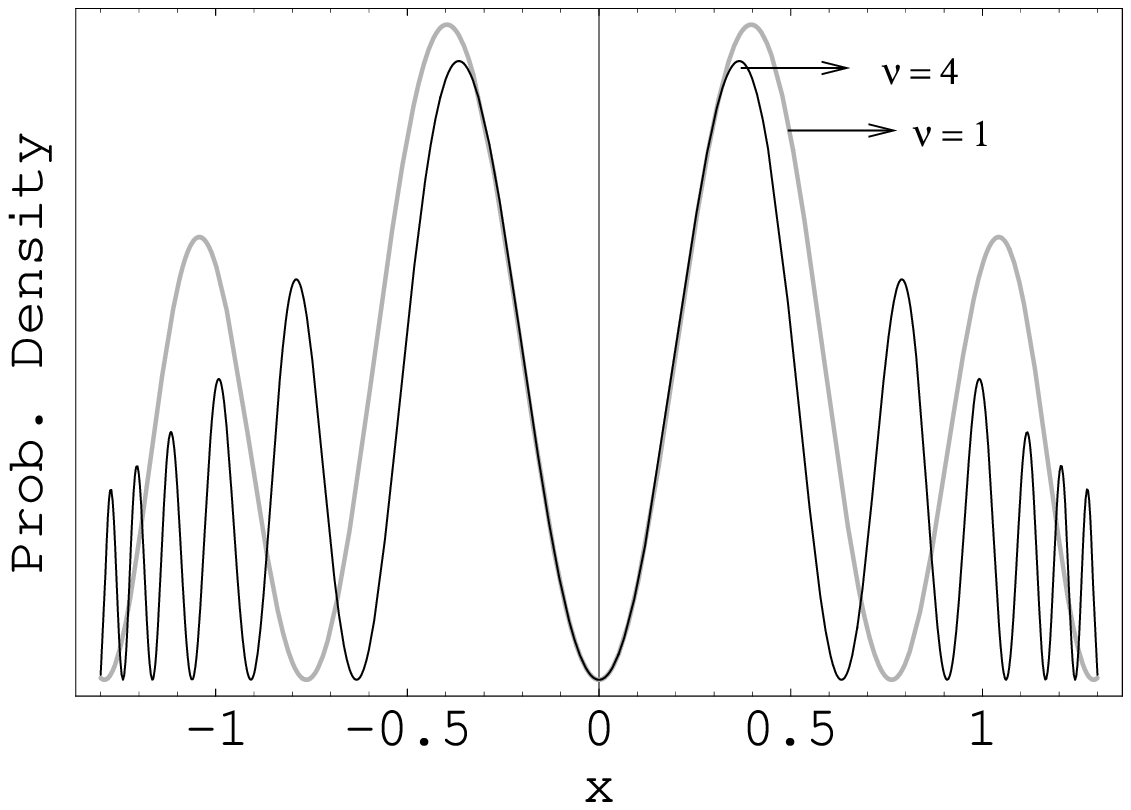}
\caption{The probability density is plotted as a function of x for
 two different values of the parameter $\nu$ = 1 and 4 for odd states.
As both the curves are showing finite area we can conclude
 the wave functions are normalizable.}
\end{figure}

The energy of the particle states can be calculated for
definite values of $\nu$ from the relation, $A_{3} = -
\frac{\nu^2}{4}$,  which in turn implies that, $E_{\nu} = -\frac{\nu^2}{8}$. 
Energy eigenvalues are negative in this case. 
The even and odd parity states are degenerate though the model
is one dimensional. This is an unusual feature which contradicts the
usual theorems on definite parity bound states in symmetric potentials. 
We have not been able to obtain a better understanding of this
curious feature of our solutions. However, the singularity in
the potentials may provide the key to the understanding of degeneracy in one
dimensions \cite{deg}.

Furthermore, one can also evaluate the expectation values of $x$ and $p$
(using standard quantum mechanics)
for these states. It turns out that:

\begin{equation}
\langle x \rangle =0 \hspace{.2in};\hspace{.2in} \langle p \rangle =0
\end{equation}
However, one can check that $\langle x^2 \rangle$ is finite, whereas
$\langle p^2 \rangle$ diverges to positive infinity. This implies that
the uncertainty $\Delta x$ is finite, but the uncertainty $\Delta p$
is infinite. The curious fact is that, the kinetic energy~
$T$ has a positively infinite expectation value, while $\langle V\rangle$
is negatively infinite but, $\langle H\rangle$ is finite. This happens
because, in the integrand, the contributions from $T$ and $V$ which make their
individual
expectation values diverge, have opposite signs and therefore, cancel!

What can we say about a bound state which is infinitely spread out in
position space? One notices that the even though the wave function is 
damped the 
number of nodes of the odd and even wave functions, within a small interval $dx$
increases with $x$. More specifically, we have

\begin{eqnarray}
2 n \pi = \sqrt{A_1} \int \left ( \cosh x \right )^{\nu} dx \nonumber \\
\frac{(2n+1)}{2} \pi = \sqrt{A_1} \int \left ( \cosh x \right )^{\nu} dx
\end{eqnarray}
for the odd and even states, respectively.
One can check easily that $\frac{dn}{dx}$ becomes progressively larger for large $x$
and fixed $dx$. This could be a reason for the divergence of the $\langle T\rangle$.
In other words, the momentum space wavefunction has a significant contribution
from the large momentum modes even though we have a localised bound state in
position space \cite{parwani}.  It may be noted that there are positive
energy bound states (in the continuum), the so--called von-Neumann--Wigner 
(vNW) states,
which do have infinite number of nodes \cite{stillinger}. However, apart
from being positive energy states, note that for the vNW states, the 
potentials are bounded below unlike the potentials we have studied here. 
A better understanding of the peculiar bound states discussed in this
article is surely desirable.


\noindent {\sf Concluding remarks :}
Let us now summarize our results. We have provided prescriptions for
constructing volcano potentials which seem to have, of late, gained relevance
in various areas of physics. After a brief discussion on classical mechanics
in a volcano potential we have studied the quantum features in some detail.
For a one--parameter family of such potentials we have been able to
write down exact bound 
state solutions for specific values of the energy and the parameters in the
potential. These solutions are normalizable in the usual sense (square
integrable) though we find degenerate states of even and odd parity.
We have shown through plots, the nature of the wave functions and
probabilities for several cases. We have also calculated expectation
values of position and momentum, the uncertainties in x and p and the
density of nodes for our wavefunctions. 

A point to note is that the potentials we have discussed lead to non-Hermitian
Hamiltonians (due to the divergence to negative infinity of the potential).
Though still a matter of debate, a fair amount of current work has
shown some of the distinctive characteristics 
(eg. real eigenvalues for complex Hamiltonians. etc.) 
of such Hamiltonians (the so--called
PT symmetric theory \cite{ptsymm}). The definition of observables as
well as the construction of quantum mechanics needs modifications
in the context of non--Hermiticity. Moreover, their physical utility
remains an open question though we have mentioned some of them here.
In fact, despite repeated use in high energy physics (braneworld models)
the aspect of non--Hermiticity has never been mentioned in any article.      

We hope the results derived in this paper may be 
useful in some way to researchers interested in pursuing the physics
of volcano potentials.

\vspace{.3cm}

SK thanks R. Parwani for discussions and some useful insights on 
various issues related to this article.
RK thanks CSIR, India for financial support through a Senior Research
Fellowship.

\maketitle

\end{document}